\documentclass[PRL,twocolumn,showpacs,groupedaddress]{revtex4-1}

\usepackage{graphicx,dcolumn,bm,}

\usepackage{pslatex}

\usepackage{amsthm}

\begin{document}
\title{Local increase of symmetry on cooling in KNi$_2$Se$_2$}

\author{James R. Neilson,$^{1,2*}$ Natalia Drichko,$^2$ Anna Llobet,$^3$ Mali Balasubramanian,$^4$ Matthew R. 
Suchomel,$^4$ Tyrel M. McQueen$^{1,2\dagger}$}
\address{$^1$Department of Chemistry, Johns Hopkins University, Baltimore, MD}
\address{$^2$Department of Physics and Astronomy and the Institute for Quantum Matter, Johns Hopkins 
University, Baltimore, MD}
\address{$^3$Lujan Neutron Scattering Center, Los Alamos National Laboratory, Los Alamos, NM}
\address{$^4$Advanced Photon Source, Argonne National Laboratory, Argonne, IL}
\address{$^*$james.neilson@colostate.edu}
\address{$^\dagger$mcqueen@jhu.edu}

\begin{abstract}   

KNi$_2$Se$_2$ exhibits an  increase of symmetry on cooling below $T\le50$~K, as observed by Raman 
spectroscopy and synchrotron x-ray diffraction.  X-ray absorption spectroscopy confirms that the symmetry 
increase is due to changes in nickel-nickel interactions and suppression of charge density wave fluctuations.  Density functional theory calculations reveal a zone-
boundary lattice instability that provides a model of the room-temperature x-ray pair distribution function data, 
but fails to describe the higher local symmetry observed for $T\le50$K.  Together, these results support many-
body correlation effects as drivers for the unusual heavy fermion electronic ground state in KNi$_2$Se$_2$.
\end{abstract}

\maketitle

Strong electron-electron and electron-phonon interactions give rise to broken symmetry states. A prototypical 
example is the charge density wave (CDW) state, which lowers the symmetry and introduces an additional periodic modulation 
of electron density.\cite{Gruner1994,Mazin08,Eiter13} CDWs belong to a larger family of spontaneous charge 
separation phenomena, such the charge ordering observed in the manganites \cite{mitchell2001spin,Bozin:2007uz} and organic conductors.\cite{PhysRevLett.87.237002}  In these materials, the formation of charge-separated states is almost always observed at low temperature, when 
thermal fluctuations are no longer able to disrupt the  electronic order. In rare situations, many-body interactions can produce a re-entrant transition in which the charge periodicity is destroyed and higher 
symmetry is restored on further cooling.\cite{PhysRevLett.87.237002,MurrayTesanovic} In fact, a decrease of 
charge-order fluctuations on cooling has been observed in metallic organic conductors with 1/4-filled bands that 
are close to charge order; the experimentally observed response is rather weak since the effect results from the electronic repulsion within a single band.\cite{PhysRevLett.96.216402}

We have recently discovered that KNi$_2$S$_2$ \cite{PhysRevB.87.045124} and KNi$_2$Se$_2$ 
\cite{Neilson_PRB_2012,Neilson_JACS_2012} have signatures of CDW fluctuations at room temperature that disappear below $T \sim 
50$~K. This surprising observation has stimulated extensive experimental \cite{PhysRevB.87.144305, 
arxiv1211.1371,  TlNi2Se2, arxiv1305.1033,KFe2As2} and theoretical \cite{MurrayTesanovic,PhysRevB.87.161122, 
Bannikov201376, 0953-8984-24-49-495501, Bannikov2013}   studies aimed at understanding the unusual nature of 
charge coupling with spin, orbital, and lattice degrees of freedom.

Most strikingly, KNi$_2$Se$_2$ shows an increase of  carrier mobility ($\mu$) below $T\sim50$~K that occurs at the same temperature at which  CDW fluctuations disappear. The 
compound enters a  heavy fermion state  below $T=20$~K with an effective electronic mass of 
$m^*\sim$ 6 to 18$m_b$, and superconductivity below $T_c$=0.8~K.\cite{Neilson_PRB_2012}  Our studies showed that the magnetic response is rather weak and temperature independent.\cite{Neilson_PRB_2012}  An explanation of a heavy fermion state was 
proposed in terms the fluctuation of a CDW involving electron-electron correlations.\cite{MurrayTesanovic}. 
The theory suggests that the loss of CDW fluctuations at low temperature results from both 
many-body physics and the hybridization between the localized, 1/4-filled band and a dispersive conduction band. 

\begin{figure}[t!]
\begin{center}
\includegraphics[width=2.8in]{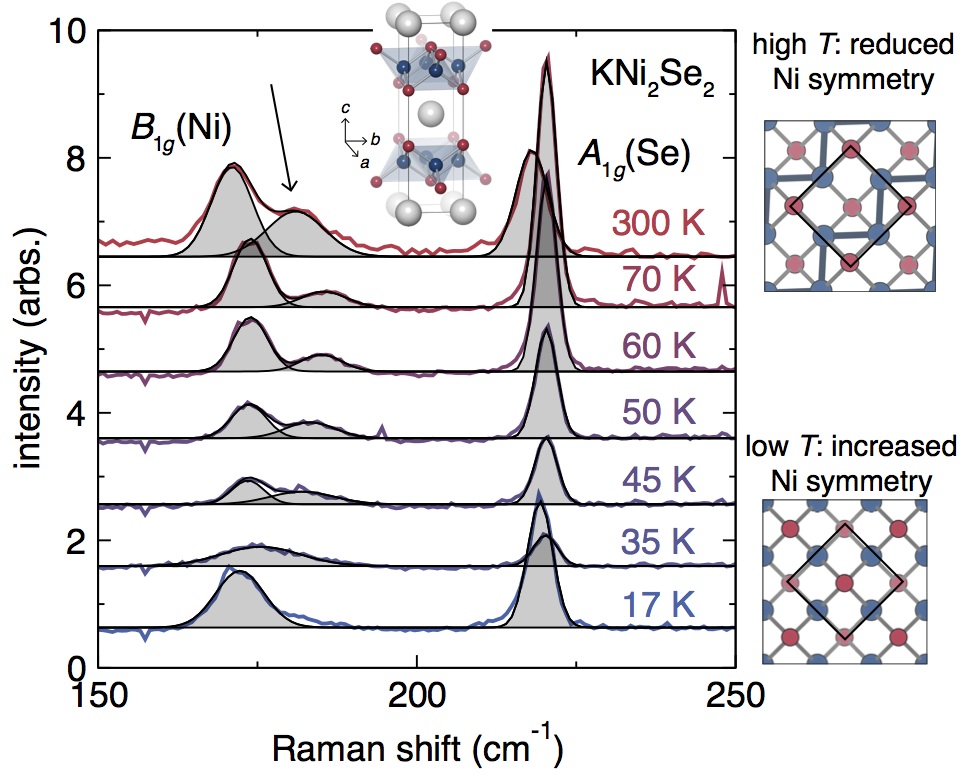}
\caption{(Color online)
 Temperature dependence of Raman spectra of KNi$_2$Se$_2$ illustrate an additional phonon mode then that 
expected from the  $I4/mmm$ spacegroup (arrow) which softens into the neighboring $B_{1g}$(Ni) mode on 
cooling. Spectra are shifted along $y$ axis for clarity.  Right: In-plane illustration of the increase of symmetry on cooling.  Top inset: the average $I4/mmm$ unit cell of KNi$_2$Se$_2$.
}
\label{fig:raman}
\end{center}
\end{figure}

In this letter, we determine that there is an increase in symmetry on cooling in KNi$_2$Se$_2$, likely driven by electronic correlations. Raman spectroscopy provides direct evidence for the increase of local point group symmetry and loss of CDW fluctuations in KNi$_2$Se$_2$ on cooling. An analysis of the structure derived from high-resolution powder synchrotron x-ray diffraction (SXRD) shows signatures of the CDW fluctuations, due to out-of-plane displacements of Ni atoms at room temperature, that then disappear on cooling. X-ray absorption spectroscopy (XAS) and the extended x-ray absorption fine structure (EXAFS) confirm that the CDW fluctuations arise from displacements of the positions of nickel atoms. Density functional theory (DFT) calculations predict an intrinsic lattice instability driven by electron-phonon coupling. A calculated imaginary frequency phonon mode at the zone boundary qualitatively reproduces the room-temperature distortion observed by synchrotron x-ray total scattering, but fails to describe the low temperature ground state of KNi$_2$Se$_2$.  In agreement with the theoretical work on the canonical CDW materials,\cite{Mazin08}, the nature of the CDW fluctuations in this material implies that the fluctuations are not driven by Fermi surface nesting, but are instead driven by local bonding interactions.

Raman spectra were obtained from the ($ab$) plane of single micro-crystals in a parallel-sided fused silica capillary (evacuated to $p_{\text{He}} = 10$~torr at RT) in a  T64000 Jobin-Yvone 
spectrometer equipped with an Olympus microscope. The laser power was below 0.3 mW to avoid sample heating, estimated by comparing Stokes and anti-Stokes spectra at $T=200$ K.  
 From a symmetry analysis of the low-temperature crystal structure presented in Ref.~\cite{Neilson_PRB_2012} (space group $I4/mmm$), we expect four 
Raman-active optical phonons.\cite{Orobengoa:ks5225}  In particular, Ni atoms in ($4d$ Wyckoff positions, Ref.~\cite{Neilson_JACS_2012})
contribute $B_{1g}$ and $E_g$ modes, of which only $B_{1g}$ is observed in the ($ab$) plane.  Se atoms ($4e$ Wyckoff 
positions, Ref.~\cite{Neilson_JACS_2012}) show $A_{1g}$ and $E_g$ Raman active modes, of which only  $A_{1g}$ is observed in the ($ab$) plane. 
Indeed, in the Raman spectra at the lowest temperatures we observe two strong bands, located at 171 and 219 cm$^{-1}$ (21.2 and 27.2 meV; Fig.~\ref{fig:raman}), which we assign to the $B_{1g}$ vibration of Ni atoms and the 
$A_{1g}$ vibration of Se atoms, respectively, based on our calculations and Ref.~\cite{PhysRevB.87.144305}.   The 
low temperature spectrum of KNi$_2$Se$_2$ has a shape very similar to the non-stoichiometric compound K$_{0.95}$Ni$_{1.86}$Se$_2$, for which no distortion is reported,\cite{PhysRevB.87.144305} even though the 
frequencies in the spectra of KNi$_2$Se$_2$ are higher.

On increasing the temperature above $T>$~50~K, the lower-frequency Raman band shifts to higher frequencies,  
and another phonon becomes well resolved at 186 cm$^{-1}$. This shape of the spectra persists to  room 
temperature.  The 
appearance of an additional phonon mode for $T>50$~K is a direct evidence for a decrease in local point symmetry of KNi$_2$Se$_2$ on warming.

\begin{figure}[h!]
\begin{center}
\includegraphics[width=3in]{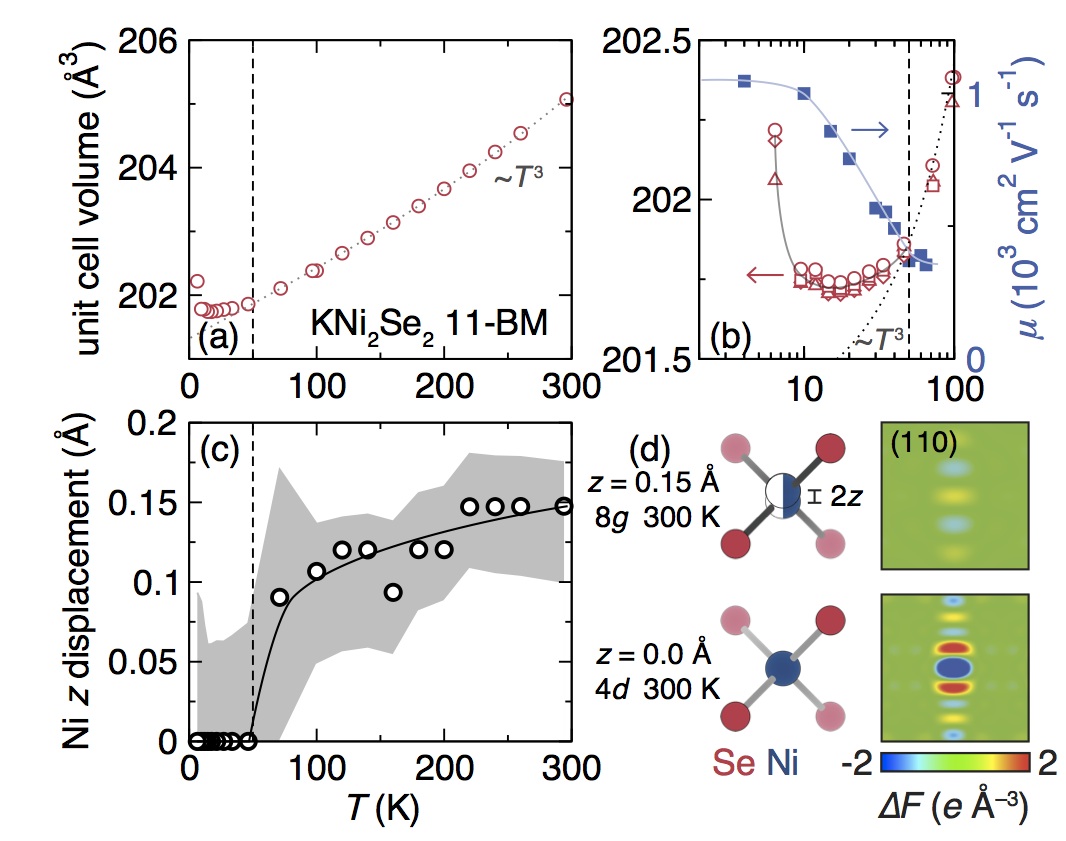} 
\caption{
(a) Unit cell volume of KNi$_2$Se$_2$ from high-resolution synchrotron powder x-ray diffraction (SPXRD, 11-BM, APS) with 
gradual change of slope near $T\sim50$~K (Dashed vertical line: $T=50$~K.  Dotted line: $ aT^3 + bT+ c$).  
(b) Negative thermal expansion below $T \sim 18$~K, measured on ($\diamond$) cooling, ($\bigtriangleup$) 
warming, ($\qedsymbol$) re-cooling, and ($\circ$) re-warming, coincides with the increase in the carrier mobility 
($\mu$, reproduced from Ref.~\cite{Neilson_PRB_2012}).
(c) The displacement of the Ni atom ($z || c$) illustrates the apparent increase in the Ni site symmetry on cooling 
near $T\sim 50$~K.  The shaded area contours a 5\% deviation from the best Rietveld g.o.f.  at each $T$.
 Solid lines guide the eye.  Error bars fall within the symbols.
 (d) Fourier difference maps from data collected at $T = 300$~K, viewed on the (110) plane, for nickel on the $4d$
(bottom) \emph{vs}.\ $8g$ (top) Wyckoff positions.}
\label{fig:symm}
\end{center}
\end{figure}

\begin{figure}[h!]
\begin{center}
\includegraphics[width=3in]{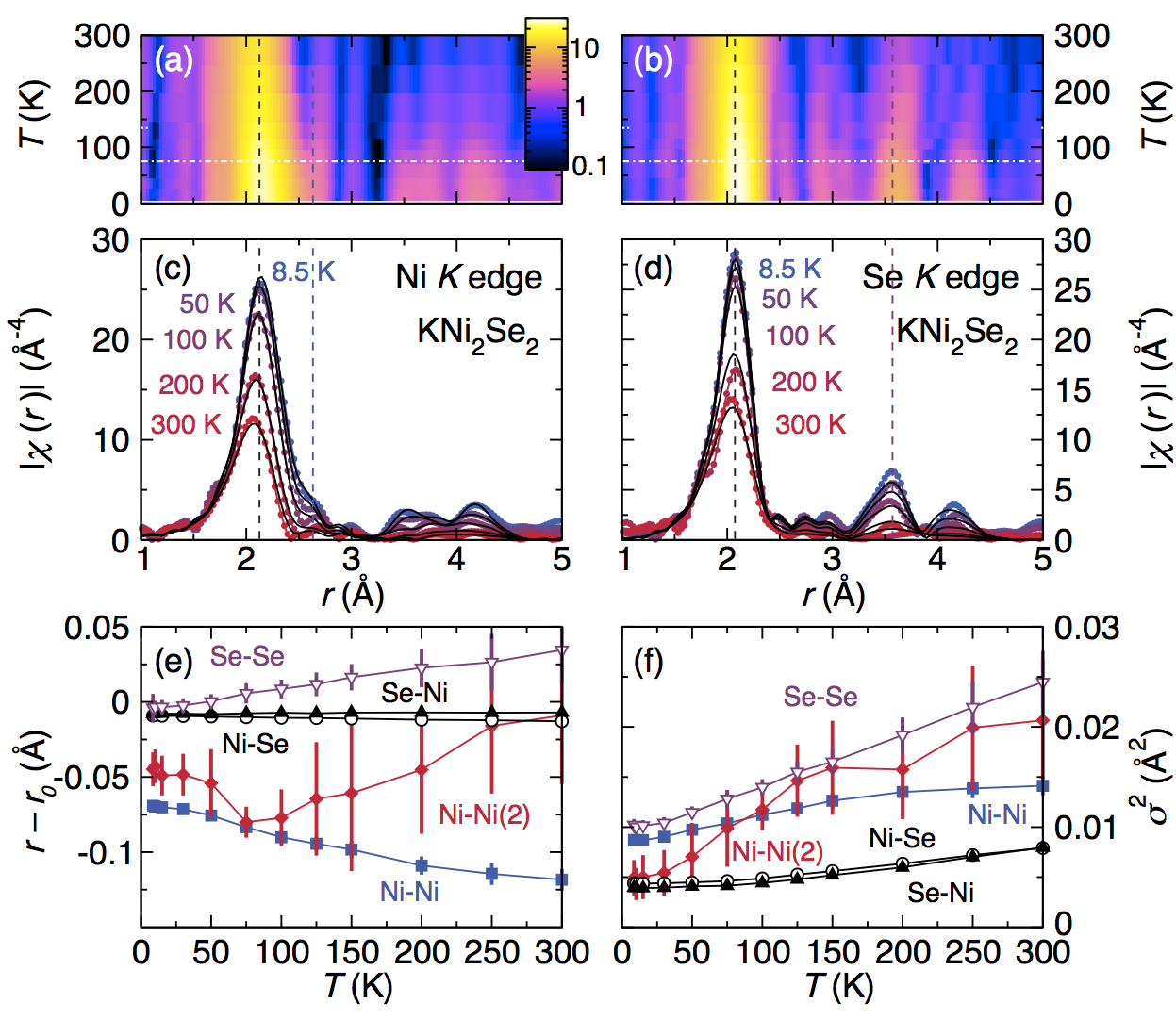}  
\caption{
Temperature dependence of the EXAFS (20-BM-B, APS) of KNi$_2$Se$_2$ measured at (a) the Ni $K$ edge and (b) the Se $K$ 
edge show a shift in spectral weight with temperature at the Ni $K$ edge predominantly around $\sim$2.6 \AA.  
Selected fits of  (c) the Ni $K$ edge and (d) the Se $K$ edge EXAFS illustrate the better quality of fit for $T\le 
75$~K.   (e) The temperature dependence of the shifts of different scattering path lengths illustrate that most of 
changes reside on the Ni sublattice [paths: Ni-Se/Se-Ni $r_0 = 2.40$~\AA; Ni-Ni $r_0 = 2.76$~\AA; Se-Se and Ni-
Ni(2) $r_0 = 3.91$~\AA].  (f) The mean squared displacements of the path lengths indicate the same trend.  Error 
bars are from the statistical error of the fit.  The vertical bars at $r\sim2.1$ \AA\ indicate the Ni-Se or Se-Ni path 
length, Ni-Ni paths at $r\sim2.6$ \AA, and Se-Se paths at $r\sim3.6$ \AA.
}
\label{fig:exafs}
\end{center}
\end{figure}

At $T = 50$~K Rietveld analysis of the high-resolution SXRD data \cite{wang:085105,PhysRevB.87.045124,gsas,expgui}  reveals a change in the average structure that gives the best 
description of the data [Fig~\ref{fig:symm}(c)].  For $T\le50~K$, the nickel atom is found at the 4$d$ Wyckoff 
position, (0, 0.5, 0.25).  However, for $T>50$~K, the goodness-of-fit (g.o.f.)  is improved by displacing the Ni atom 
from this ideal position to the $8g$ Wyckoff position $(0,0.5,z)$ with a 50\% fractional occupancy, as illustrated by 
the $T=300$~K Fourier difference maps in Fig.~\ref{fig:symm}(d). This demonstrates an apparent decrease in the  
local symmetry of Ni positions on heating above $T>50$~K. In fact, if we describe the Ni atoms using the $8g$ Wyckoff positions in a $I4/mmm$ unit cell, we would expect two Ni Raman-active 
modes ($A_{1g}$ and $B_{1g}$) in the ($ab$) plane, in agreement with our experimental Raman spectra at $T>50$~K. 

XAS data (beamline 20-BM-B \cite{Ravel:ph5155,Newville:ph5149}) within the measured temperature range from $T$=300 to 8.5 K show no significant changes of the formal 
valence (Ni$^{1.5+}$, Se$^{2-}$), as neither Ni nor Se $K$ edges move in energy within the uncertainty of the XAS 
experiment ($\Delta E \sim 0.05$~eV).  This confirms our previous speculation that the charge fluctuations are not due to single-ion effects, but instead result from nickel-nickel interactions.  However, several different methods indicate  slight changes in the unit cell 
and details of crystal structure as a function of temperature, showing that the local symmetry change is due to the 
Ni atoms.

There is  a significant redistribution of the Fourier-transformed EXAFS signal (in real space) for the Ni $K$ edge as a function of temperature [Fig.\ref{fig:exafs}(a)], indicative of a change in the Ni coordination environment. In contrast, there is no shift in the Se $K$ edge EXAFS [Fig.\ref{fig:exafs}(b)].  The EXAFS confirms that it is the nickel atoms that are responsible for the apparent local increase of symmetry on cooling.

The EXAFS data were quantitatively modeled using the average $I4/mmm$ crystal structure to simulate the 
path lengths out to a cluster size of $\sim4.5$~\AA.  The  quality of fit of the model, as judged by the $R$-factor ($R = [\sum ($data$ - $fit$)^2]/[\sum($data$)^2$]),  is 
temperature independent  from $8.5 \le T \le 75 K$ ($\sim 0.7\pm0.1\%$); above that temperature the $R$-factor 
increases nearly linearly to $2.2\%$ by $T=300$~K.  This reduced fit quality is also visible 
in Fig.\ref{fig:exafs}(c) and (d) and indicates that the average crystal structure does not accurately describe the 
local Ni and Se environments for $T > 75$~K.   Attempts to fit a distorted model to the higher temperature EXAFS 
were unreliable. The fits of the ideal undistorted structure provide physically meaningful temperature dependences of the  path length changes ($r-r_0$) and 
mean-squared displacements ($\sigma^2$).

The Ni-Se (Ni edge, Ni to Se path; 4$\times$ degenerate) and Se-Ni (Se edge, Se to Ni path; 4$\times$ degenerate)  
paths show very little temperature dependence [Fig.\ref{fig:exafs}(e)] and the $\sigma^2$ exhibit the expected 
gradual increase on warming. The Se-Se path (Se edge, Se to Se nearest neighbor) also does 
not show any temperature dependent inflections. In contrast,  the Ni-Ni path (the 4$\times$ degenerate 
nearest neighbor nickel separation) has a significant  temperature dependence, showing that the 
distance between  Ni atoms decreases on cooling from $T=300$~K,  saturating around $T\sim50$~K, which is 
consistent with the Ni atom position from Rietveld analysis.  The next-nearest Ni-Ni separation [Ni-Ni(2); 4$\times$ degenerate] shows a sharp inflection at $T=75$~K, consistent with the out-of-plane displacement of the Ni position determined by X-ray diffraction.  While the significant error bars and increase in $
\sigma^2$ prevent an accurate assignment of the exact disposition of that path length at high temperatures, it is 
clear that the microscopic Ni environment changes at $T\sim75~K$.

\begin{figure}[t!]
\begin{center}
\includegraphics[width=3in]{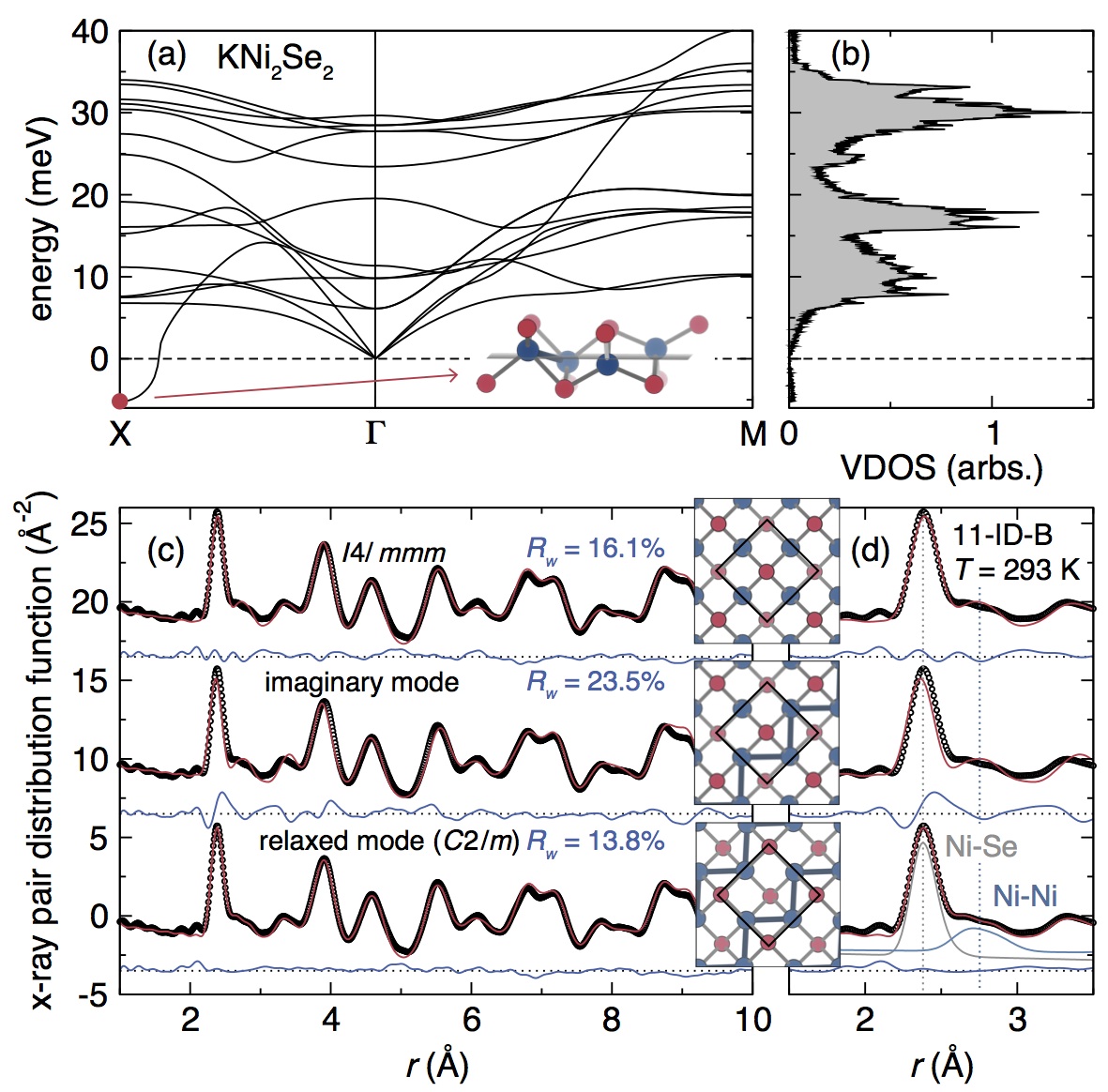}
\caption{(Color online)
(a) DFT calculated phonon dispersion curves with a negative energy mode [$\vec{q} = (\frac{1}
{2}~0~0)$]; the illustrated mode corresponds to out-of-plane Ni displacements [Ni: blue (dark) circles; Se: red 
(light) circles; amplitude exaggerated for clarity].
(b) Calculated vibrational densities of states (VDOS).
(c) X-ray pair distribution function (PDF; 11-ID-B, APS) of KNi$_2$Se$_2$ ($T = 300$~K; filled circles; reproduced and offset for 
clarity).  Lines through the data are the calculated PDF from the $I4/mmm$ crystal structure (top), the imaginary 
frequency eigenvector from the DFT phonon calculation (middle), and its least-squares relaxation within space 
group symmetry $C2/m$ (bottom); differences are below.
(d) Illustration of Ni-Se and Ni-Ni pairwise correlations.  Insets illustrate the corresponding [Ni$_2$Se$_2$] plane 
(bonds denote nearest neighbors). }
\label{fig:dftpdf}
\end{center}
\end{figure}

DFT calculations were carried out to gain insight to the nature of the observed structural distortions.  
A linear-response DFT calculation of the zone center phonons  \cite{elk,PhysRevLett.100.136406}  for the low-temperature $I4/mmm$ crystal structure of KNi$_2$Se$_2$ generally agrees with the results of the Raman scattering measurements \footnote{see EPAPS for calculation details and mode energies}, as well as with the mode energies and assignments reported in Ref.~\cite{PhysRevB.87.144305}.  Calculation of the in-plane phonon dispersion reveals an imaginary frequency eigenvector at the $X$ point in the 
Brillouin zone [$\vec{q} = (\frac{1}{2}~0~0)$, $E = 5.2i$ meV, Fig.~\ref{fig:dftpdf}].  Imaginary frequency modes 
from DFT can be indicators  of lattice instabilities, as exemplified by the localized, aperiodic distortions in Bi
$_2$Ti$_2$O$_7$.\cite{Shoemaker_PRB_2010}  In the present case, this prediction for a distorted ground state 
contrasts with the experimentally observed undistorted ground state at the lowest temperatures measured  (see 
Figures~\ref{fig:symm} and \ref{fig:exafs}, and Ref.~\cite{Neilson_PRB_2012}).  This contradiction suggests that the 
level of theory used for the DFT calculations lacks the ability to adequately describe the structure. However, this 
predicted lattice instability  does resemble the distortion of KNi$_2$Se$_2$ observed above $T\sim50$~K.  The 
primary polarization components of the imaginary frequency eigenvector are out-of-plane Ni displacements, as 
illustrated in Fig.~\ref{fig:dftpdf}(a) [$\pm z$, with a small component along $x$, such that $z\sim3x$].

To determine the validity of the distortion predicted by DFT, X-ray total scattering experiments were performed to identify the distortion (APS, 11-ID-B, 90 ke, $Q_{max}$= 30 \AA$^{-1}$ \cite{Chupas_2003,fit2d,PDFgetX2,PDFgui})
The DFT-predicted distortion (after scaling the amplitude by a least-squares fit) does not directly improve the fit to the x-ray PDF at room temperature where we see the evidence for CDW [Fig.\ref{fig:dftpdf}
(c) and (d); \footnote{See EPAPS, Table~II}].   However, using {\sc isodistort} \cite{isodisplace}, 
the imaginary-frequency zone-boundary phonon polarizations were decomposed into the subgroup symmetry, 
$C2/m$ \footnote{See EPAPS, Table II}, which allowed for an excellent description to the PDF  out to $r=10$~\AA\ after a least-squares refinement [Fig.
\ref{fig:dftpdf}(c) and (d)].   The resulting configuration contains short and long Ni-Ni bonds, akin to the long-range ordered CDW modulations in the structurally related compounds, SrRh$_2$As$_2$\cite{PhysRevB.85.014109} and KCu$_2$Se$_2$.\cite{PhysRevB.67.134105}
While the bond distances from the room temperature PDF can be described by a $C2/m$ distortion, the overall 
crystal symmetry determined by high-resolution SXRD remains $I4/mmm$. While the space group setting that we used with $C2/m$ should have more modes visible in the Raman spectrum, the true atomistic configuration may possess pseudo-symmetry, such that the local point group symmetries do not provide additional modes than beyond what is observed.  The Ni atoms exhibit the most significant distortions.  Therefore, we expect that only the Ni point group symmetry 
is significantly changed with temperature, and 
the distortions are oriented out of the plane. These results are in agreement with the  Rietveld analysis of SXRD 
data [Fig.\ref{fig:symm}(c)]and Raman scattering.

An additional piece of evidence for the importance of electronic correlations is an increase in unit cell volume on  
cooling below $T\sim20$~K,  yielding a negative coefficient of thermal expansion [Fig.~\ref{fig:symm}(a,b)].  This result is 
highly reproducible on several heating and cooling cycles, akin to KNi$_2$S$_2$.\cite{PhysRevB.87.045124}   A negative coefficient of thermal expansion can be a signature of the  
formation of a heavy fermion state (confer CeAl$_3$ \cite{PhysRevLett.35.1779}). The 
observed temperature dependence of the unit cell coincides with changes in the carrier mobility and suggests that 
the structure and electronic properties are highly correlated.

In conclusion, this letter shows a direct evidence of a decrease of local symmetry in  KNi$_2$Se$_2$  
with increasing  temperature.  Out-of plane Ni atom displacements are a signature of CDW fluctuations driven by local Ni-Ni interactions, rather than Fermi surface nesting, that do not exhibit any long range order, and do not reduce the average $I4/mmm$ symmetry of the unit 
cell. Raman spectroscopy directly demonstrates the lowering of local symmetry at high temperatures.   Generally, a raising of symmetry on cooling is unusual, as it naively suggests an increase in configurational 
entropy, when considering only the degeneracy of lattice modes. 
Hybridization of localized orbitals with the conduction band, as previously proposed,\cite{MurrayTesanovic} may provide a state in which the total entropy is decreased by such a raising of lattice symmetry and explain why our DFT calculations fail to correctly predict the ground state of this material.   Such interactions between localized and delocalized electronic systems have been proposed as an origin of nematicity,  coherence/incoherence crossover and heavy electrons in the iron-based superconductors,\cite{KFe2As2,PhysRevB.87.161122,PhysRevB.64.085109,PhysRevLett.87.237002,yin2012correlation,Caron_2011,Caron_2012} . The work presented here unambiguously demonstrates the presence and importance of charge-lattice coupling in ThCr$_2$Si$_2$-type materials.

We thank James Murray and Zlatko Tesanovic for inspiration and motivation, as well as Igor Mazin for helpful 
discussions and Kevin Bayer and Karena Chapman for data collection at 11-ID-B.  This research and the Raman spectrometer are principally 
supported by the US Department of Energy (DoE), Office of Science, Office of Basic Energy Sciences (BES), 
Division of Materials Sciences and Engineering under Award DE-FG02-08ER46544, as well as  start-up funds from 
the Johns Hopkins University and funding from the David and Lucile Packard Foundation.  This research has 
benefited from the use of beamlines 11-BM-B, 11-ID-B, and 20-BM-B at the Advanced Photon Source at Argonne 
National Laboratory, supported by the U.S. DoE, Office of Science, Office of Basic Energy Sciences, under 
Contract No. DE-AC02-06CH11357.

\bibliography{../../../Neilson_References}

\end{document}